\documentclass[twocolumn,amssymb,footinbib, nobibnotes, aps, reprint]{revtex4-2}
\pdfoutput=1 
\usepackage{graphicx}
\graphicspath{{./}}
\usepackage{bm}
\usepackage{braket} 
\usepackage{physics, mathtools}
\usepackage{hyperref}
\hypersetup{colorlinks,
	urlcolor=[rgb]{0,0.6,0.6},
	linkcolor=[rgb]{0,0.0,0.8}}

\renewcommand{\vec}{\mathbf}
\newcommand{\mat}[1]{\mathbf{#1}}

\begin{document}

\title{Branched flows in active random walks and the formation of ant trail patterns}

\author{King Hang Mok}
\affiliation{Max Planck Institute for Dynamics and Self-Organization (MPI-DS), 37077 G\"ottingen, Germany}
\affiliation{Faculty of Physics, Georg-August-Universit\"at G\"ottingen, 37077 G{\"o}ttingen, Germany}
\author{Ragnar Fleischmann}
\email{ragnar.fleischmann@ds.mpg.de}
\affiliation{Max Planck Institute for Dynamics and Self-Organization (MPI-DS), 37077 G\"ottingen, Germany}

\date{\today}

\begin{abstract}
\noindent \textit{Branched flow} governs the transition from ballistic to diffusive motion of waves and conservative particle flows in spatially correlated random or complex environments. It occurs in many physical systems from micrometer to interstellar scales. In living matter systems, however, this transport regime is usually suppressed by dissipation and noise. 
In this article we demonstrate that, nonetheless, noisy active random walks, characterizing many living systems like foraging animals, and chemotactic bacteria, can show a regime of branched flow. To this aim we model the dynamics of trail forming ants and use it to derive a scaling theory of branched flows in active random walks in random bias fields in the presence of noise. We also show how trail patterns, formed by the interaction of ants by depositing pheromones along their trajectories, can be understood as a consequence of branched flow.  

\end{abstract}

\maketitle

\section{Introduction}\label{sec:intro}
\begin{figure}[ht!]
	\centering
	\includegraphics[width = \columnwidth]{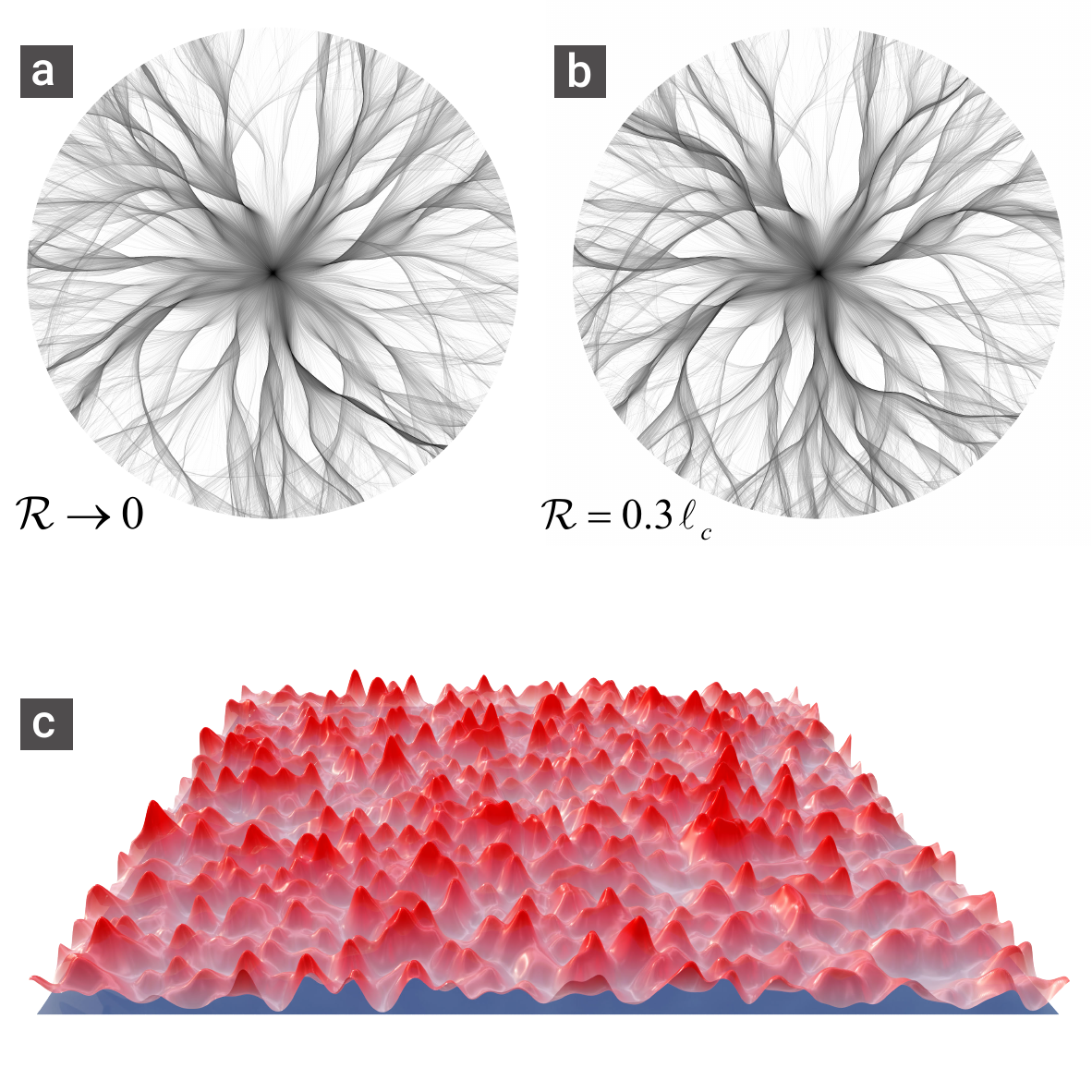}
	\caption{\textbf{Branched flow of ant trajectories.} Example densities of model trajectories of ants exiting from a hole in the center of a disc with a random pheromone field and leaving the disc at the edge. Comparison of the dynamics according to (\textbf{a}) the noise free ($\gamma_i=0$) differential equations of motion Eq.~\ref{eq:eom} with  (\textbf{b}) the integro-differential dynamics of Eq.~\ref{eq:eomkernel} with a detection radius of 30\% of the correlation length of the pheromone field.  (\textbf{c}) Example of a random pheromone field model with density $n=1.0$. }
	\label{fig:BF}
\end{figure}

\noindent
High intensity fluctuation patterns and extreme events are hallmarks of branched flow, which very generically occurs in the propagation of waves, rays or particles in weakly refracting correlated random (or even periodic) media~\cite{Heller2021,Kaplan2002,Metzger2010,Metzger2014,Daza2021}. It is a ubiquitous phenomenon and has been observed in many physical systems, e.g. in electronic currents refracted by weak impurities in high mobility semiconductors~\cite{Topinka2001,Aidala2007}, light diverted by slight variations of the refractive index~\cite{Patsyk2020,Patsyk2022}, microwaves propagating in disordered cavities~\cite{Hoehmann2010,Barkhofen2013}, sound waves deflected in the turbulent atmosphere~\cite{BlancBenon2002,Ehrhardt2013} or by density fluctuations in the oceans~\cite{Wolfson2001}. Wind driven sea waves are piled up by eddies in the ocean currents to form rogue waves~\cite{White1998,Heller2008,Ying2011,Ying2012,Green2019} and tsunamis are focused to a multiple of their intensity even by minute changes in the ocean depth~\cite{Degueldre2016}. Branched flow dominates propagation on length scales between the correlation length of the medium and the mean free path of the flow that is traversing it, i.e. a regime between ballistic and diffusive transport in an environment with \textit{frozen} or \textit{quenched} disorder. An example of a branched flow is shown in Fig.~\ref{fig:BF}a.

Experiments that have analysed the laws of motion of individual Argentine ants (Linepithema humile) during trail formation~\cite{Perna2012} have inspired us to study if branched flow can also occur in random walks in biology. In living systems, motion in general is overdamped and the phase space structures responsible for branched flows can not form. However, frequently the inevitable input of energy happens in form of self-propulsion and motion is best described as a \textit{active random walk} (often referred to as \textit{active brownian particles})~\cite{Berg1984,Schweitzer2007,Romanczuk2012,Tailleur2022}. And in many situations these biological or biologically inspired active random walks are subject to \textit{bias fields}, like for instance the distribution of food in animal foraging or bacterial chemotaxis but also of chemicals acting as fuel for self-propelled colloids~\cite{Hokmabad2022,Golestanian2022,Saha2014,Duan2021}. In the dynamics of ants, the bias field is representing pheromones left by other ants along their paths. Using this example we show that in active random walks in correlated random bias fields one can observe the same phenomenology of branched flow as in conservative flows, and that due to the heavy-tailed density fluctuations associated with branched flow, this can have severe implications on pattern-formation.

The agents in active random walks, however, are usually not only influenced by the bias fields but their directionality and/or position are also subject to temporal fluctuations (which we will assume to be uncorrelated in the following), leading also to diffusion. We study how this \textit{stochastic diffusion} destroys branched flow, by deriving a universal scaling theory as the main result of this paper.

Finally, we simulate trail formation by explicitly  modelling the pheromone deposition along ant trajectories and its feedback on the trajectories of following ants,  resembling the phenomenology of the experimental observations of Ref.~\cite{Perna2012} and demonstrate its connection to the phase space structures of branched flows.

\section{Ant dynamics}\label{sec:dynamics}
\noindent
The complex collective and social behaviour of ants is of course governed by many ways of interaction: visual, tactile and chemical. Here, following the lead of Ref.~\cite{Perna2012}, we want to concentrate on a simple model of the dynamics of Argentine ants as they are influenced by the pheromones deposited by other ants. Please note that, due to several uncertainties in the experimental knowledge that would require us to make assumptions in the model and fit many parameters, our aim will not be to quantitatively reproduce experimental findings but to formulate a clean, simplified model that captures the main aspects observed qualitatively and phenomenologically and which allows us to study the fundamental implications of correlated bias fields.

\begin{figure}[b]
	\centering
	\includegraphics[width = 0.9\columnwidth,clip=true]{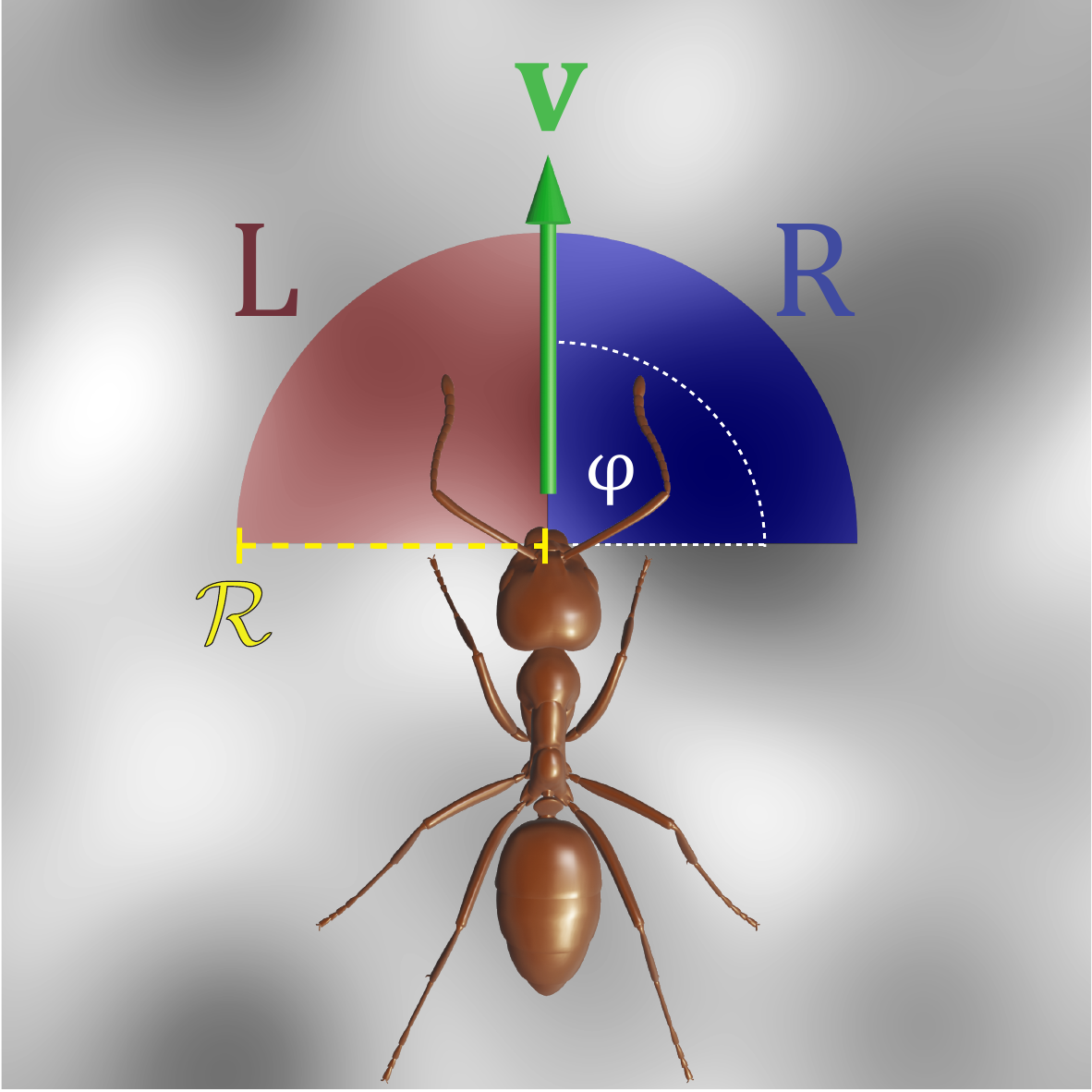}
	\caption{\textbf{Pheromone detection}. An ant running with velocity $\vec{v}=(0,v)^T$ in y-direction (i.e. $\varphi=\pi/2$). The ant is assumed to integrate the pheromone concentration field (grey scale in the background) over quarter-circular domains of radius $\cal{R}$ to the left (L,red) and right (R, blue) of its path and on average change its heading according to Eq.~\ref{eq:weber}.}
	\label{fig:pheromone}
\end{figure}

The experiments reveal that the (mean) change in the direction of the ants' trajectories due to their perception of spatial variations in the pheromone concentration can well be described by a (generalized)\textit{ Weber's law}. This is a fundamental law of sensory psychophysics describing perception that does not depend on the total  sensory input but only on its relative changes, see e.g.~\cite{Johnson2002,Zwislocki2009}.  Here the directional change $\Delta \varphi$ of the ant's velocity $\vec{v}=(v \cos \varphi, v \sin \varphi)^T$ in a given time interval $\Delta t$ was found to be
 \begin{equation}
	\left\langle \Delta \varphi\right\rangle  =A_{\Delta t}\, \frac{L-R}{L+R+T_0},
	 \label{eq:weber}
\end{equation}
where  $\left\langle. \right\rangle$ denotes an appropriate ensemble average (because $\Delta \varphi$ itself is a stochastic quantity, as will be discussed later). The proportionality constant $A_{\Delta t}$ depends on the time interval. The quantities $L$ and $R$ are measuring the concentration $c(\vec{r})$ of pheromones that the ant detects, integrated over certain domains to the left and right of its projected path, respectively, as illustrated in Fig.~\ref{fig:pheromone} ($\vec{r}=(x,y)^T$ is the two dimensional position vector). At very low pheromone concentrations the detection threshold of the ants' sensors will eventually be reached and therefore Weber's law was  
adjusted by a threshold parameter $T_0$.

In the limit $\Delta t\rightarrow 0$, Eq.~\ref{eq:weber} becomes an integro-differential equation as described in Appendix~\ref{app:eom}. To facilitate simulating many realizations of random bias fields and large ensembles of random walkers, we simplify these integro-differential equations into ordinary differential equations by Taylor-expanding the concentration field and taking the limit $\mathcal{R}\rightarrow 0$ (see Appendix~\ref{app:eom}). As hinted earlier, the changes of the direction of the ants are not deterministic, but stochastic quantities. Therefore we are also introducing noise terms to the equation of motion, finally yielding stochastic differential equations:
\begin{equation}
	\begin{aligned}
		dx &=\cos \varphi\,dt+\gamma_1\,dW_x \\
		dy&=\sin \varphi\,dt + \gamma_1\,dW_y\\
		d\varphi &= \alpha \, \frac{\grad c \cdot \vec{n}}{c + T_0} \,dt + \gamma_2 \,dW_\varphi.
	\end{aligned}
	\label{eq:eom}
\end{equation}
Here $\vec{n}=(-\sin \varphi,\cos \varphi)^T$ is the unit normal vector of the ant's velocity and the proportionality constant $\alpha$ is a measure of the sensitivity of the ant's response to spatial variations in the pheromone field. The quantities $dW_i$ are independent white noise (Wiener) processes and the parameters $\gamma_1=\sqrt{2\,D_1}$ and $\gamma_2=\sqrt{2\,D_2}$ quantify the strength of translational and rotational diffusion, respectively~\footnote{To solve these equations of motion numerically we have both used a standard stochastic Runge-Kutta scheme as well as the automated solver choice provided by the julia package \textit{DifferentialEquations.jl} with compatible results.}. In the case of the ant dynamics we will always assume $\gamma_1=0$, but since our results will also be interesting and valid for other active random walks like bacterial chemotaxis, we will later also let $\gamma_1>0$. Figure~\ref{fig:BF} illustrates that the (noise free) dynamics of these equations well captures the essentials of the dynamics obtained using the kernel method (described in Appendix~\ref{app:eom}), here shown in a random pheromone field model which will be introduced in the next section.

\begin{figure}[bt!]
	\centering
	\includegraphics[width = \columnwidth]{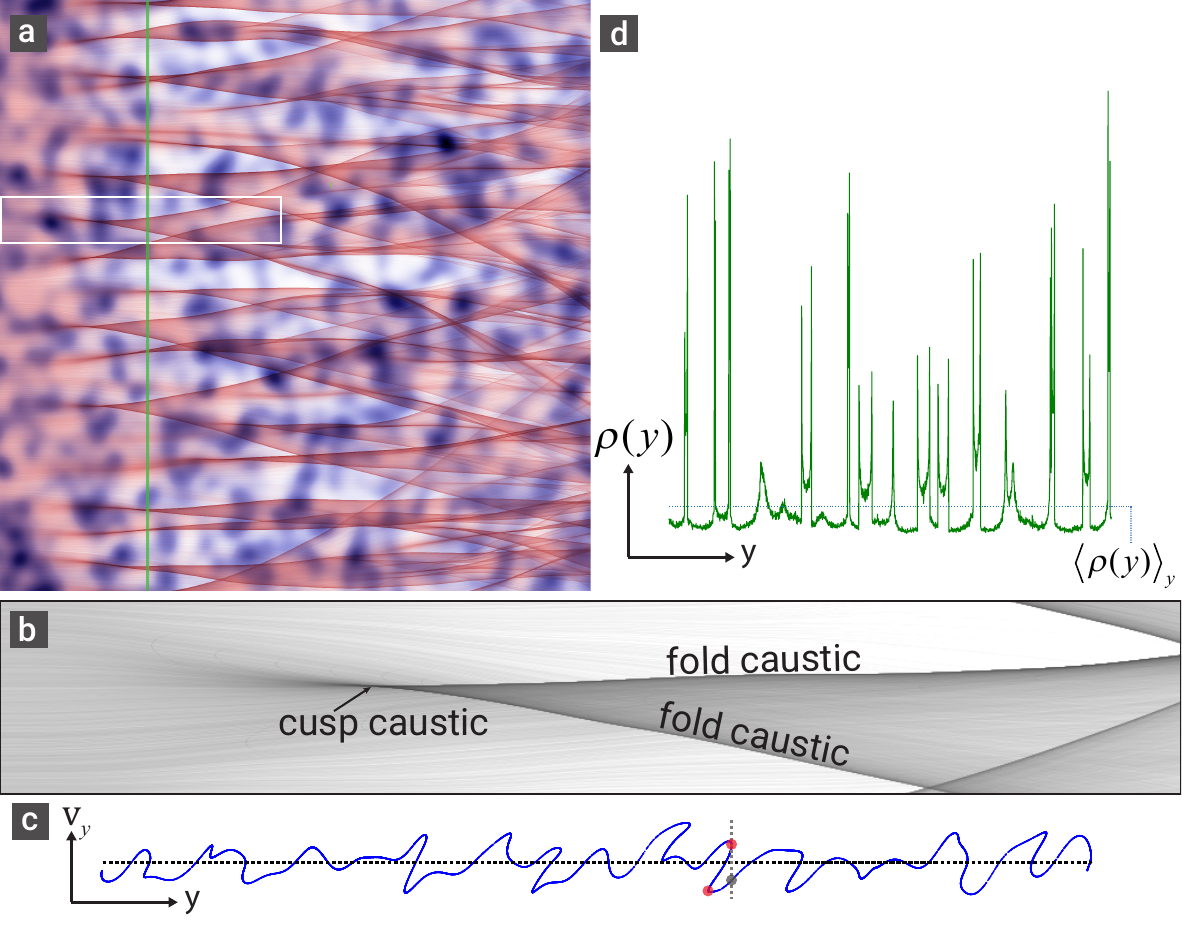}
	\caption{\textbf{Branched flow of initially parallel trajectories.} \textbf{(a)} The density (red) of a bundle of trajectories starting on the left of the shown region homogeniously distributed on the y-axis with initially parallel horizontal velocity $\vec{v}(0)=(1,0)^T$, propagating over a random pheromone field (indicated in a white-to-blue colormap in the background). \textbf{(b)} Zoom in on the density in the white-rimmed box in (a). It illustrates how a branch is initially formed by the occurrence of a random \textit{cusp caustic} from which two \textit{fold caustics} spawn. 
		\textbf{(c)} Poincare surface of section along the green line in (a). The manifold representing the initially parallel ray bundle has evolved into a contorted line. Caustics are points with vertical tangent. The red dots indicate the fold caustics from (b). The grey dot indicates the second solution at the same spatial coordinates as the upper fold caustic.
		\textbf{(d)} Cut through the density $\rho(y)=\rho(x=\mathrm{const.},y)$ along the green line in (a), showing the density singularities associated with the caustics and indicating how branched flow leads to heavy-tailed density fluctuations.}
	\label{fig:pw1}
\end{figure}

\section{Dynamics in random fields and typical length scales}\label{sec:random}
\noindent
In many situations the environment of an active random walk is complex and the bias field will be best described as a \textit{correlated random field}. In the ant dynamics this might e.g.\ be in a crowded environment, where many ant trajectories cross paths. In the experiments of Ref.~\cite{Perna2012}, where ants exit a hole in the centre of a disc in random directions, the estimated pheromone field does not show signs of trail formation in the first 10 minutes, but appears to be random. In branched flows even minute (but correlated) random variations in the environment lead to heavy-tailed, branch-like density fluctuations in the flow. We can therefore hypothesize that these structures can be the seeds of emerging patterns like the formation of ant-trails. We will therefore begin our analysis of ant dynamics by studying motion in random fields.

We will use a very simple random field model with only few parameters: imagine a concentration field that is created by randomly placed Gaussian-shaped mono-disperse pheromone "droplets" (see Appendix~\ref{app:rf}). It can be characterized by a correlation length $\ell_c$ and the density $n$ of droplets per area $\ell_c^2$. Its mean and variance are then given by $\left\langle c \right\rangle = n/\ell_c $ and $\sigma_c^2=n/(\pi \ell_c^4)$. An example for $n=1$ is shown in Fig.~\ref{fig:BF}c.

For analysing the interplay of branched flow and stochastic diffusion we will need to know the scaling of the typical length scale of branched flows governed by the equations of motion Eq.~\ref{eq:eom}. 

One prominent characteristic of branched flows is the occurrences of \textit{random caustics}~\cite{Berry1980,Kulkarny1982,White1984,Klyatskin1993, Kaplan2002, Heller2021}. Caustics, first studied in ray optics, are contour-lines or -surfaces in coordinate space on which the number of solutions passing through each point in space changes abruptly. They are also singularities in the ray (or trajectory) density. Figure~\ref{fig:pw1} illustrates the connection of branched flows and caustics.

For the dynamics in correlated random potentials it is well established that the typical length scale of branched flow is given by the average distance ($d_c$) a ray or trajectory has to propagate until it reaches the first caustic~\cite{Kaplan2002,Metzger2010,Barkhofen2013,Metzger2014}. More details can be found in Appendix~\ref{app:scaling}, where we transfer the scaling arguments derived earlier to the ant dynamics. The essence of these arguments is that in a paraxial approximation $d_c$ is reached if the diffusion due to the random bias field in direction perpendicular to the initial propagation direction covers a correlation length of the random field, i.e.
(assuming initial motion in $x$ direction)
\begin{equation}
	\left<(y-y_0)^2\right> \approx\;\ell_c^2,
	\label{eq:scalingArg}
\end{equation}
where $<\cdot>$ is the average over many random fields of equal characteristics or, due to self-averaging, the average over initial conditions in sufficiently large systems. From this we find (see Appendix~\ref{app:scaling}) that 
\begin{equation}
	d_c\propto \left( \frac{\ell_c}{\alpha r_c}\right)^{2/3},
	\label{eq:dcscaling}
\end{equation}
with the \textit{correlation radius} of the random force
\begin{equation}
	r_c= \left( \int_{-\infty}^{\infty} C_F(x,0) \;dx\right)^{1/2} 
	\label{eq:corrrad}
\end{equation}
where $C_F(x,y)=\left<F_y(x,y)\,F_y(0,0)\right>/\alpha^2$ is the correlation function of the force $F_y=\left(\alpha \partial_y c\right) / \left( c+T_0\right)$.
In the following we evaluated $r_c$ numerically.

\begin{figure}[tb!]
	\centering
	\includegraphics[width=\columnwidth]{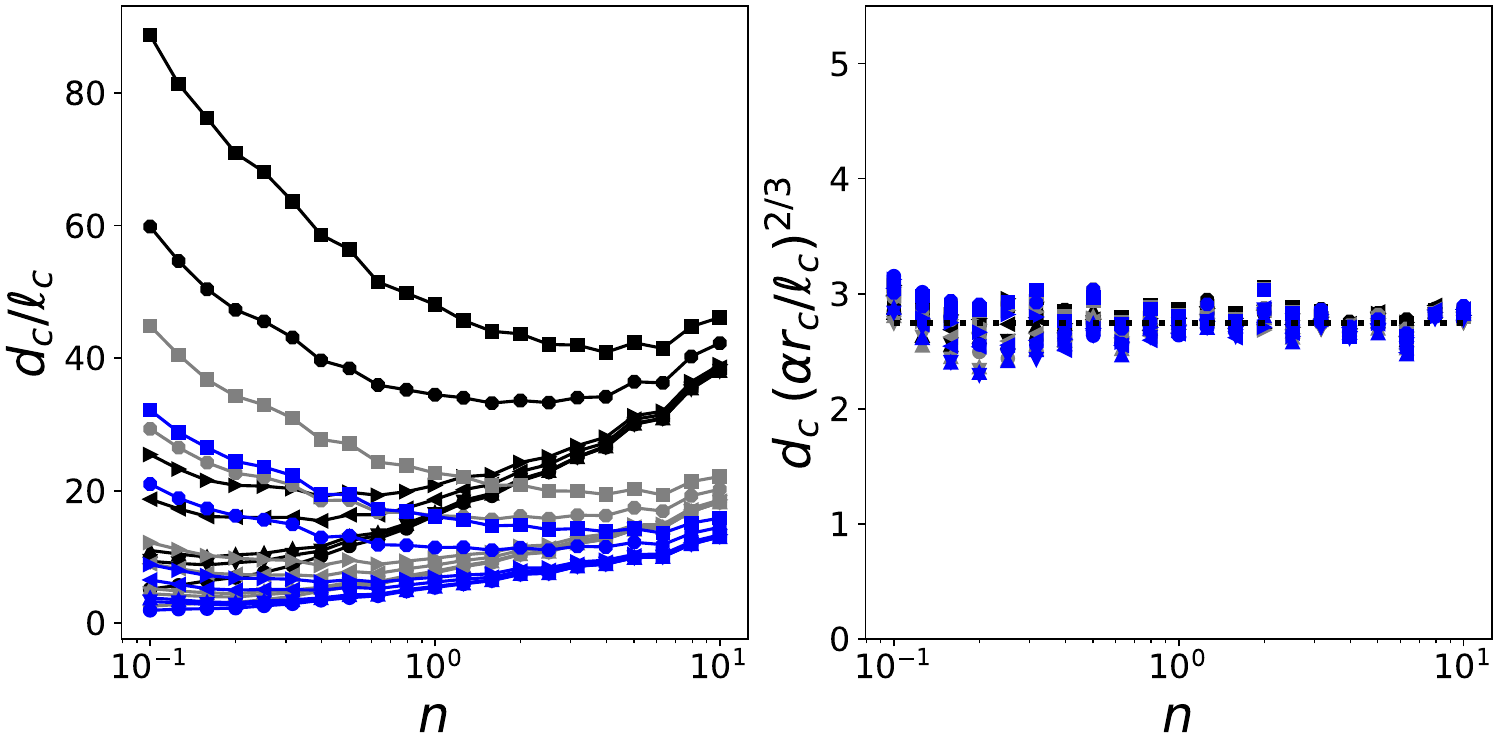}
	\caption{\textbf{Characteristic length scale.} The numerically obtained \textit{mean distances to the first caustic} $d_c$ for random pheromone fields for $441$ combinations of different parameters  $n$, $\alpha=\left[0.1, 0.3, 0.5\right]$ and $T_0=\left[0.0, 0.032, 0.064, 0.32, 0.64, 3.2, 6.4\right]/\ell_c^2$ plotted versus $n$ (the \textit{trivial} parameters are fixed to the values $\ell_c=0.014$ and $w=1$.). Each datapoint is averaged over 10 realizations of the pheromone field and $10\,000$ initial conditions for each field. The simulations clearly confirm the scaling derived in Eq.~\ref{eq:dcscaling} by data collapse: \textbf{(Left)} $d_c$ in units of $\ell_c$. \textbf{(Right)} The same data as in the left panel scaled by the right hand side of Eq.~\ref{eq:dcscaling}. To make is easier to follow individual curves they are coloured according to the $\alpha$ values: black (0.1), grey (0.3) and blue (0.5). } 
	\label{fig:Dc}
\end{figure}

To confirm the scaling behaviour of Eq.~\ref{eq:dcscaling} we obtained the \textit{first caustic statistics} for a range of different parameters of the random pheromone fields by numerically integrating the stability matrix along trajectories as described in Appendix~\ref{app:stabMat}.
The results are shown in Fig.~\ref{fig:Dc}.

\section{Diffusion and branched flow}\label{sec:diffusion}

\begin{figure}[b!]
	\centering
	\includegraphics[width = \columnwidth]{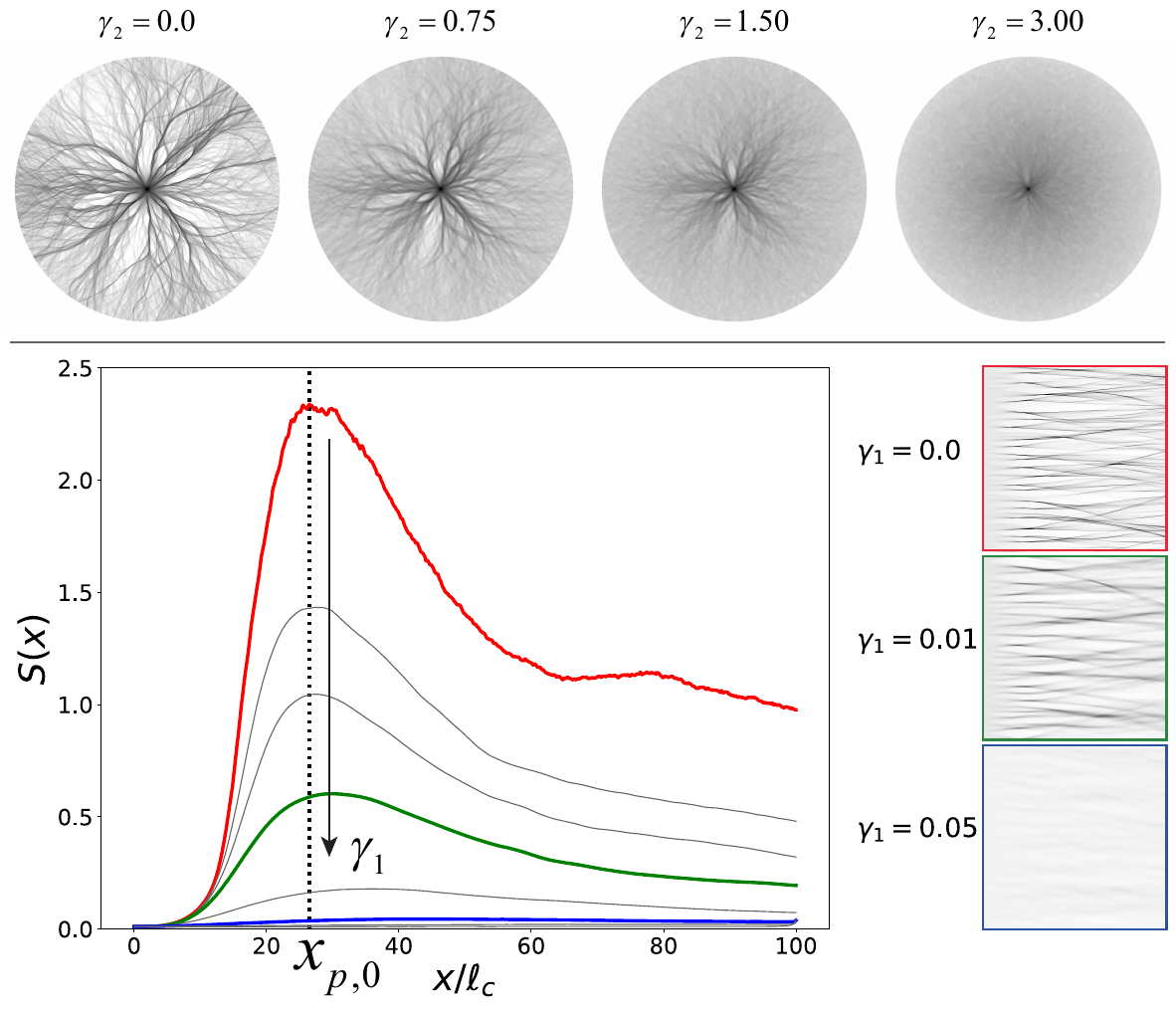}
	\caption{\textbf{Impact of stochastic diffusion.} Illustration of the suppression of branched flow by increasing stochastic diffusion.
	\textbf{(Upper panel)} Flow from a point source with increasing stochastic rotational diffusion (growing $\gamma_2$). The other parameters are $\alpha=0.2$, $T_0=0.0$, $\ell_c=0.005$, and $n=1.0$. The system size is set to unity $\mathcal{L}=1$. See also \textit{Video 1} in the \hyperref[app:videos]{ supplementary material}. \textbf{(Lower panel)} Scintillation index for initially parallel branched flows for increasing translational stochastic diffusion ($\gamma_1=0.0,\cdots,0.15$). The scintillation index is averaged over $y$ and 60 realisations of the fields. The three insets on the right illustrate single realizations corresponding to three of the scintillation index curves (as indicated by color).  The other parameters are $\alpha=0.025$, $T_0=0.0$, $\ell_c=0.01$, and $n=1.0$.}		
	\label{fig:Scintillation}
\end{figure}

\begin{figure}[tb!]
	\centering
	\includegraphics[width = \columnwidth]{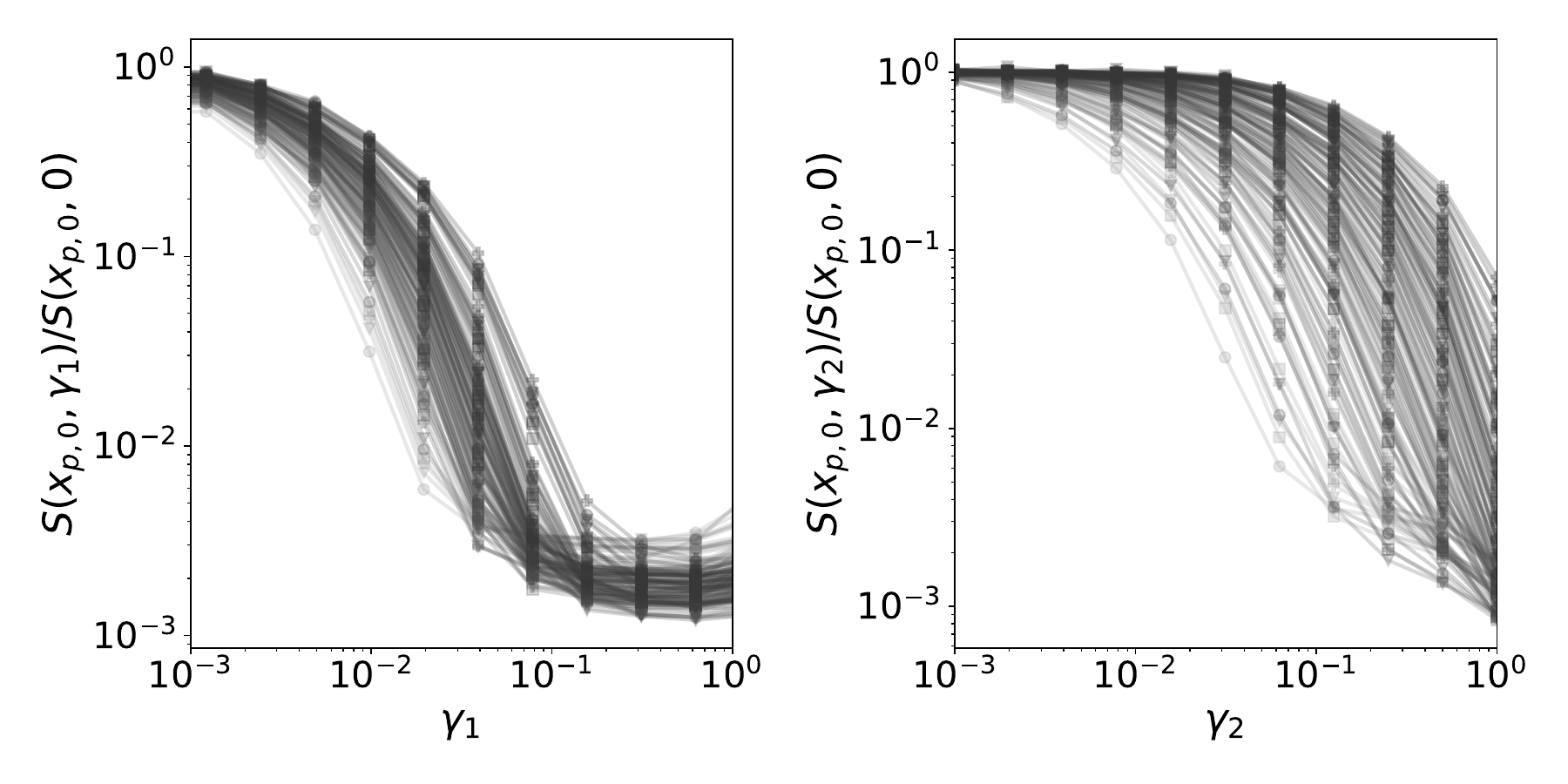}
	\caption{\textbf{Suppression of branched flow by diffusion}.
		The scintillation index $S(x_{p,0})$ at the position $x_{p,0}$ of the peak (of height $S_0(x_{p,0})$ ) in the noise free branched flow as a function of the fluctuation strength ($\gamma_i$) for \textbf{(left panel)} translational diffusion and \textbf{(right panel)} rotational diffusion for 160 combinations of the other parameters: $\alpha=[0.01, 0.02, 0.04, 0.08,0.16]$, $n=[1,2,4,8]$, $T_0=[0,0.1,1.0,10.0]/\ell_c^2$, and $\ell_c=[0.005,0.01]$ (shown are 126 parameter combinations for which $x_{p,0}$ was within the range $[0.05,0.8]$). In each simulation the density is estimated using $300\,000$ trajectories, and the scintillation index is averaged over 60 realizations of the random pheromone field.  }
	\label{fig:ScintUnscaled}
\end{figure}

We are now equipped to study how the stochastic diffusion terms in Eq.~\ref{eq:eom} will suppress branched flows, which is illustrated in Fig.~\ref{fig:Scintillation}. To quantify the suppression we use the so-called \textit{scintillation index}  of the trajectory density
\begin{equation}
	S(x)=\frac{\left\langle \rho^2(x,y) \right\rangle - \left\langle \rho(x,y) \right\rangle^2}{\left\langle \rho(x,y) \right\rangle^2},
\end{equation}
where $x$ is the main propagation direction of the flow and the average $\left\langle \cdot \right\rangle$ is taken over realizations of random fields (and in practice, to save computation time, also over the direction perpendicular to the propagation direction). For simplicity we will use initially parallel flows in $x$ direction in the following (instead of point sources where the main propagation direction would be radial). The region of the strongest branches in a branched flow is visible as a pronounced peak in the scintillation index (cf.\ Refs.~\cite{Barkhofen2013,Metzger2014}). We will use the value of the scintillation index at the propagation distance where the branched flow is most pronounced in the absence of stochastic diffusion, i.e.\ at the peak position $x_{p,0}$ for $\gamma_i=0$ (as indicated in the lower panel of Fig.~\ref{fig:Scintillation}). Please note, that to be able to compare the trajectory densities (and thus the scintillation index) for different values of the stochastic diffusion terms, we need to make sure that a well defined state, i.e.\ a \textit{non-equilibrium steady state} (NESS), has been reached. Particles (= ants) are entering from the left and exit to the right (and on the left) of the integration region (for practical reasons we are using periodic boundaries in $y$ direction). With increasing stochastic terms the total integration time until all particle have left the integration region is increasing. Figure~\ref{fig:ScintUnscaled} shows $S(x_{p,0})$ normalized to the peak height $S_0(x_{p,0})$ of the scintillation index in the absence of stochastic terms. 

We argue that branched flow will be suppressed when the stochastic terms are interfering with the basic mechanism of caustic creation. That means, when, at the mean time to the first caustic, the stochastic terms cause a diffusive standard deviation in $y$ of the same order of magnitude as the diffusion due to the (static) pheromone field, i.e.\ proportional to the correlation length $\ell_c$ of the random field.
For translational diffusion this scaling argument for the characteristic fluctuation strength $\gamma^*_1$ reads 
\begin{equation*}
	\left<(y-y_0)^2\right> = {\gamma^*_1}^2 \; d_c \propto \ell_c^2 
\end{equation*}
and thus by using Eq.~\ref{eq:dcscaling} we can define
\begin{equation}
	\gamma^*_1= \left(\alpha\;\ell^2_c \; r_c\right)^{1/3} 
\label{eq:gamma1}
\end{equation}

\begin{figure}[tbh!]
	\centering
	\includegraphics[width = \columnwidth]{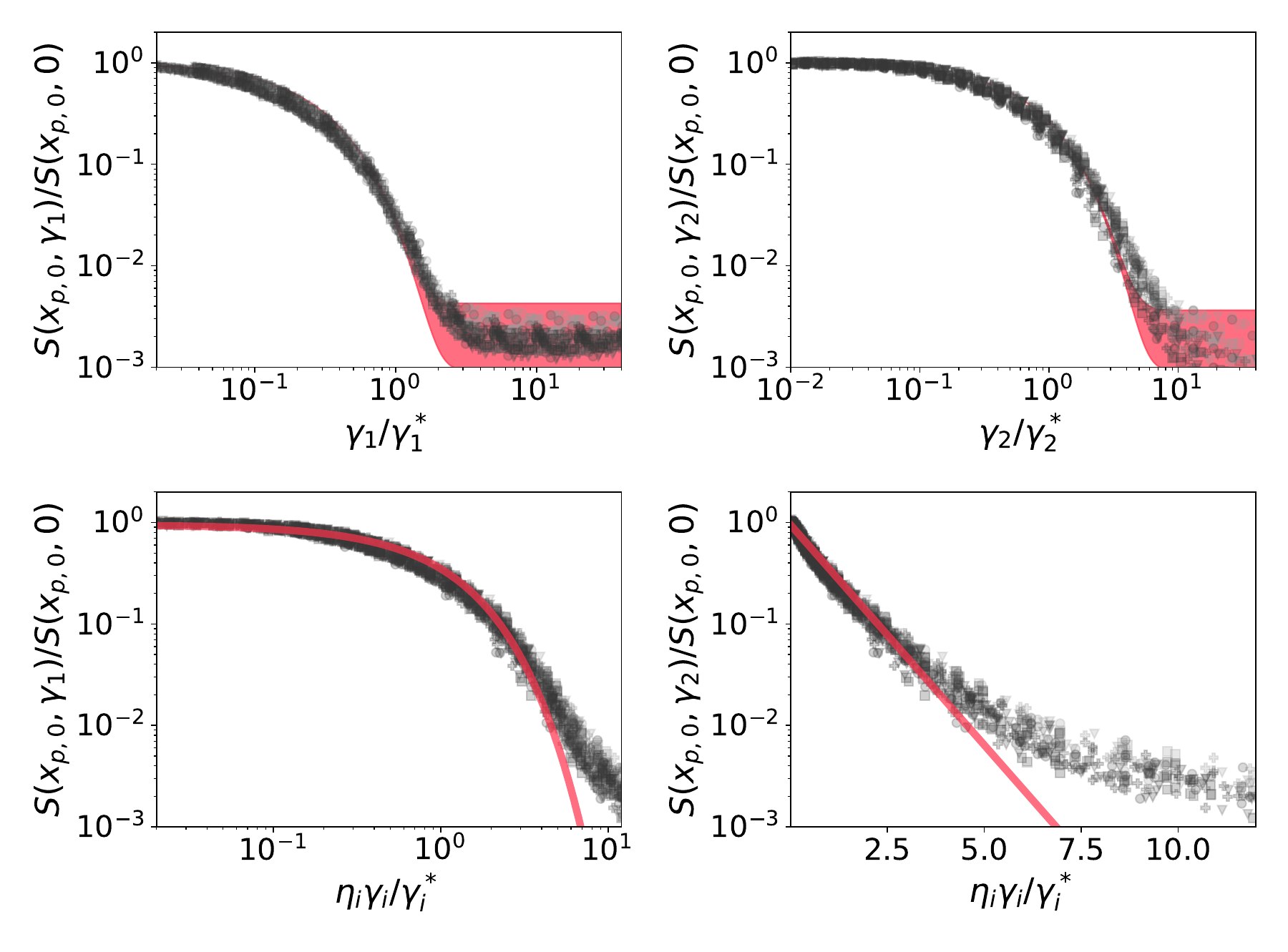}
	\caption{\textbf{Scaling theory.} The upper panels show the same data as in Fig.~\ref{fig:ScintUnscaled} but with abscissas scaled by $\gamma^*_1$ \textbf{(upper left)} and $\gamma^*_2$ \textbf{(upper right)}, respectively. The lower panels show the same data plotted together scaled according to Eq.~\ref{eq:joined} on a double logarithmic scale \textbf{(lower left)} and on a semilogarithmic scale \textbf{(lower right).} The \textbf{red} curves in the lower panels are the exponential function $y=\exp(x)$ demonstrating that the inital suppression of the scintillation index is exponential. The red shaded regions in the upper panels illustrate that the further progression of the curves can be understood by the residual scintillation in the trajectory density due to the finite number of trajectories used in the simulation. The shaded areas are defined by $S(x_{p,0})/S_0(x_{p,0})= \exp\left({-\eta_i\gamma_i/\gamma^*_i}\right) + S_\mathrm{sim}/S_0(x_{p,0})$, where $S_\mathrm{sim}\approx2.9\cdot10^{-3}$ is the numerically observed average residual scintillation index (at large $\gamma_i$). (Note that the width of the shaded area is thus caused by the spread of $S_0(x_{p,0})$ for the different parameters).}
	
	\label{fig:ScintScale}
\end{figure}

For rotational diffusion finding $\gamma^*_2$ is easy since in the paraxial approximation the pheromone field and the stochastic term enter the equations of motion on the same footing, i.e.\  without further calculation we can simply assume $\gamma^*_2\propto\gamma_p$ (cf.\ Appendix~\ref{app:scaling}) and we thus define
\begin{equation}
	\gamma^*_2=\alpha\; r_c.
	\label{eq:gamma2}
\end{equation}

\begin{figure*}[bht!]
	\centering
	\includegraphics[width = \textwidth]{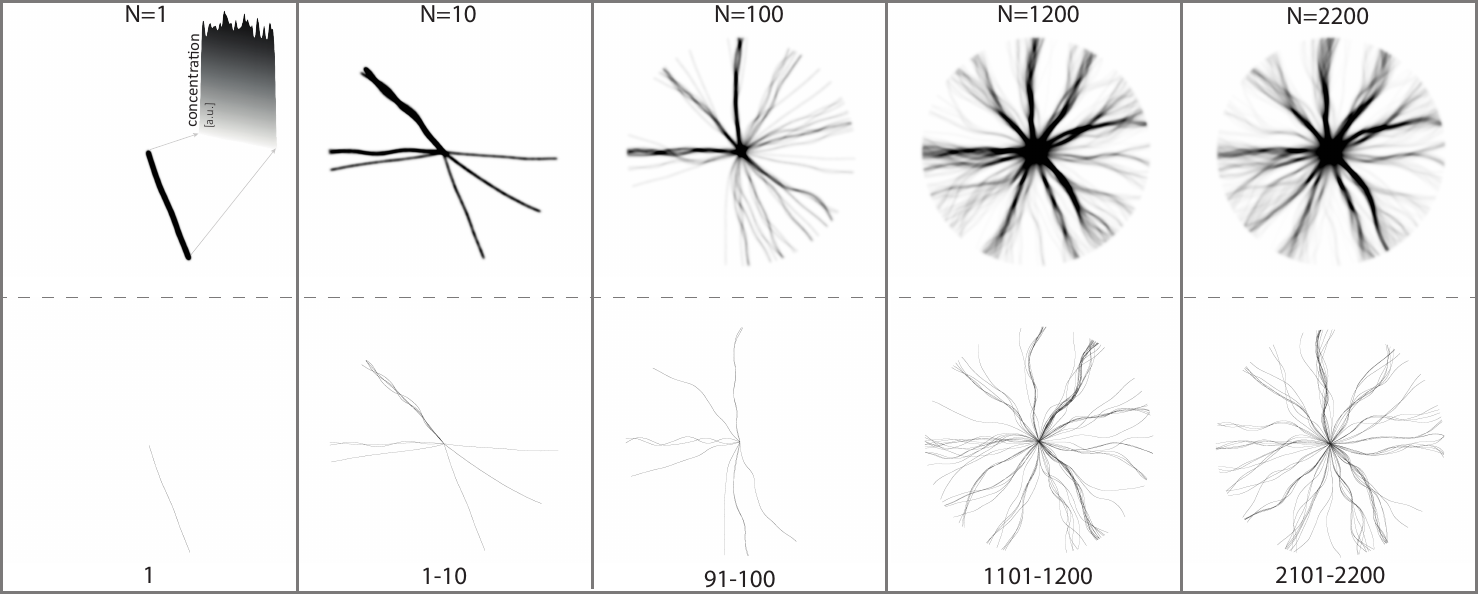}
	\caption{\textbf{Trail pattern formation.} In the \textbf{upper row} the cumulated pheromone field (in an initially pheromone free circular arena) after a number of $N$ ants have transversed the arena form the centre to the boundary with random initial direction. In the \textbf{lower row} the trajectories of the first ant entering the arena (leftmost panel) and $10$, respectively $100$, trajectories of consecutive ants at later times. (A  more detailed description of the model and the parameters used are given in the text.)}
	\label{fig:trails}
\end{figure*}

Using Eqs.~\ref{eq:gamma1} and \ref{eq:gamma2} to rescale the abscissas of the data from Fig.~\ref{fig:ScintUnscaled}, we find excellent data collapse in the upper panels of Fig.~\ref{fig:ScintScale}, confirming our scaling argument for the suppression of branched flows. 

The functional form of this suppression of branched flow, however, is surprising. One might naively expect that the Gaussian propagator of diffusion also leads to a Gaussian suppression of the scintillation index (this is e.g.\ what we would observe for the scintillation index of a periodic stripe pattern if we'd convolve it with a Gaussian kernel). In contrast we observe that the suppression is exponential in $\gamma_i$, and not in $\gamma_i^2$. This can be seen in the lower panels of Fig.~\ref{fig:ScintScale}. We find that for two orders of magnitude in the scintillation index we can write 
\begin{equation}
	S(x_{p,0}) =S_0(x_{p,0}) \exp\left({-\eta_i\frac{\gamma_i}{\gamma^*_i}}\right),
	\label{eq:joined}
\end{equation}
with constants $\eta_1\approx3.70$ and $\eta_2\approx 1.32$.

\section{Trail formation}\label{sec:trails}
\noindent
Finally, we are going to study the dynamics of our model ants interacting with each other by depositing pheromones along their trajectories. As motivated earlier, we will not be aiming at a quantitative comparison with the experiment. Instead we want to restrict ourselves to observing the basic phenomenology and to illuminate the phase space structures connected with trail formation. 

As before, in our model, ants are entering from a point source in the centre of a circular environment and are taken out of the system, when they reach the boundary. Since in the experiment of Ref.~\cite{Perna2012} the ants tend to remain at the boundary of the circular arena once they have reached it, this appears to be an adequate approximation that allows for a clear model setup and well defined states. The model ants are simulated one by one, each entering the arena in a random initial direction, following Eq.~\ref{eq:eom} in the pheromone field deposited by their predecessors, and depositing pheromones themselves along their path. After deposition, pheromones will diffuse and evaporate on slower time scales. Again, following Ref.~\cite{Perna2012}, we will assume that these time scales are long enough so that we can neglect these effects. In the following we model the deposition to be in droplets of fluctuating quantity along the path (fluctuating uniformly in an interval $[0,2\,\overline{c}]$) at times  $t_j=j \Delta t_d$ which get smeared out by a Gaussian $\exp[-(\vec{r}-\vec{r(t_j)})^2/\sigma_0^2]/(\pi \sigma_0^2)$, with arbitrary mean $\overline{c}$, $\sigma_0=0.005 L$ (where $L$ is the diameter of the arena), and $\Delta t_d$ such that the concentration fluctuates by approximately $10\%$ along the path, as illustrated in the upper leftmost panel of Fig.~\ref{fig:trails}. The remaining parameters are chosen to be $\alpha=0.2$, $T_0=0.005$, $\gamma_1=0$ and $\gamma_2=0.2$.

Figure~\ref{fig:trails} exemplifies the evolution of the pheromone field and the trail pattern in our model setup. The phenomenology is similar to that observed in the experiment. Trails start to form but might shift or depopulate over time while new trails are forming. Figure~\ref{fig:phasespace} and Video 2 in the \hyperref[app:videos]{supplementary material} illustrate the connection of the observed trail patterns to branched flows by showing the densities of \textit{potential trajectories}: after $N$ ants have deposited pheromones along their trajectories the possible trajectories of ant No. $N+1$ in the current pheromone field have been calculated and their density plotted (for each denstiy shown we simulated 200000 trajectories). To make the phase space structures that are developing clearer, in addition we have plotted the potential trajectory density in the absence of stochastic diffusion, i.e.\ for $\gamma_1=\gamma_2=0$. We clearly see, that after the initial approximately one hundred trajectories the phase space structures resemble those of a branched flow in a random environment and the most pronounced branches with their caustics correspond to the trails that have developed and are continuing to form. 

\begin{figure*}[th!]
	\centering
	\includegraphics[width = \textwidth]{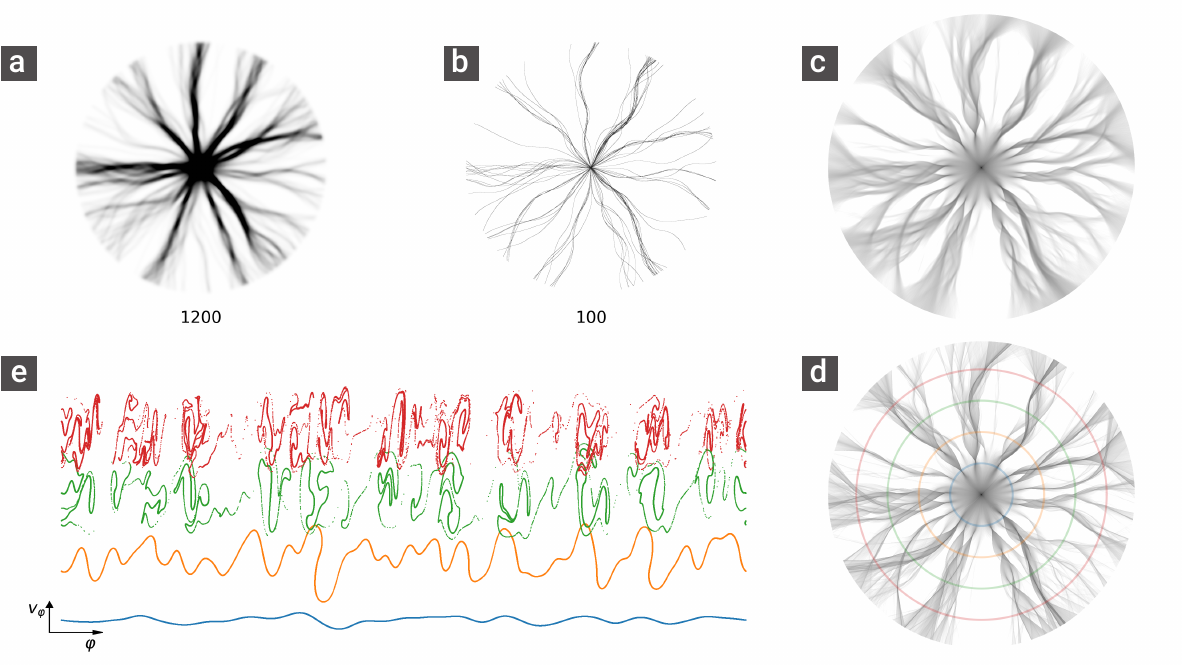}
	\caption{\textbf{Phase space analysis} of the evolving trail pattern (the time evolution is shown in Video 2 in the supplemental material~\cite{Note2}). In the \textbf{upper row}, panel \textbf{(a)} shows the accumulated pheromone field deposited by $1200$ ant trajectories. Panel \textbf{(b)} shows the latest $100$ ant trajectories. Panel \textbf{(c)} reveals the density (on a logarithmic scale) of \textit{potential trajectories} the next ant could follow in the current pheromone field. Panel \textbf{(d)} in the \textbf{lower row} shows the density (on a logarithmic scale) of \textit{potential trajectories} in the absence of stochastic diffusion terms. Panel \textbf{(e)} shows Poincaré surfaces of section (PSSs) along the circles of corresponding colour indicated in the right panel. In each PSS the angular velocity $v_\varphi=\dot{\varphi}$ is plotted against the polar coordinate angle $\varphi$ at which the trajectory is intersecting the corresponding circle.}
	\label{fig:phasespace}
\end{figure*}

\section{Conclusion}
We have demonstrated that active random walks in correlated bias fields, even in the presence of dissipation and stochastic forces, can show a regime of branched flow. We have derived a scaling theory to estimate the strength of the stochastic forces that still allow for branched flows to form. Since it is a very robust mechanism creating density fluctuations with heavy-tailed distributions and extreme events, branched flow can be crucial in the selection of random dynamical patterns forming in transport of active matter on length scales between ballistic and diffusive spread. We have exemplified this in a simple model of the trail formation of pheromone depositing ants. 

\begin{acknowledgments}
RF would like to thank Vito Trianni for making him aware of Ref.~\cite{Perna2012} and thus triggering this project.
\end{acknowledgments}

\appendix 
\section{Equations of motion}
\label{app:eom}
\begin{figure}[b]
	\centering
	\includegraphics[width = 0.9\columnwidth,clip=true]{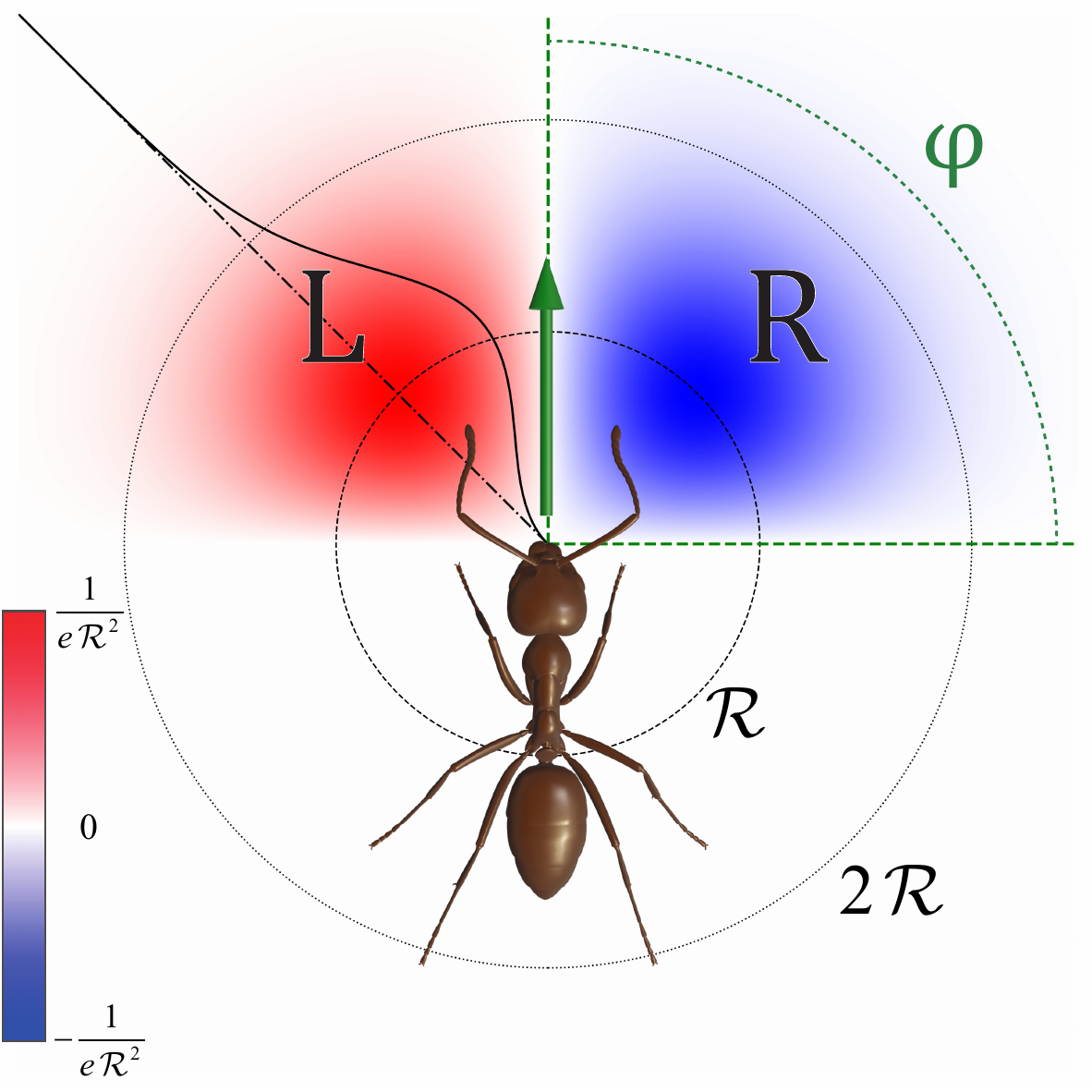}
	\caption{\textbf{Integral kernel.}  The kernel function of Eq.~\ref{eq:kernel} is shown in colour code (blue to red) for an ant running in y-direction, i.e. $\varphi=\pi/2$. The solid black curve shows the value of the kernel along the dashed dotted diagonal cut. The maximum is reached at a distance  $\mathcal{R}$ from the origin (i.e. the position of the ant).}
	\label{fig:Kernel}
\end{figure}

\noindent
We have slightly generalized the model of Ref.~\cite{Perna2012} (which we have summarized in the paragraph containing Eq.~\ref{eq:weber})  by introducing a smooth kernel function $K(r')$ instead of sharp domains. It is defined such that
\begin{align}
	L-R &=\iint_{\mathbb{R}^2} K\left(\mathbf{D}_{-\varphi}(\vec{r}'-\vec{r})\right)\;c(\vec{r'}) \, dx' \, dy' 	\label{eq:kernalints1} \\
	L+R &=\iint_{\mathbb{R}^2}  \left| K\left(\mathbf{D}_{-\varphi}(\vec{r}'-\vec{r})\right) \right|  c(\vec{r'}) \,dx' \, dy', 
	\label{eq:kernalints2}
\end{align}
when the ant is positioned in $\vec{r}$ and running in direction $\varphi$. $D_{\theta} =\left(\begin{array}{cc}
	\cos \theta & -\sin \theta\\
	\sin \theta & \cos \theta
\end{array} \right)$  denotes the rotation matrix. The integral of $\left|K\right|$ is normalized to 1. Even though a smooth kernel is more realistic than sharp domains, we don't know its actual shape  and thus have to assume some form. In the following for simplicity we chose a kernel based on a two times differentiated Gaussian, i.e.
\begin{equation}
	K(x,y)=2\,x\,y\, e^{-(x^2+y^2)/\mathcal{R}^2} \Theta(x) /\mathcal{R}^4,
	\label{eq:kernel}
\end{equation}
where $\Theta(x)$ is the Heaviside step function. The extent of the area over which the ant can detect the pheromone is parameterized by $\mathcal{R}$.  The kernel is illustrated in Fig.~\ref{fig:Kernel}. 

We can now write the equations of motion as integro-differential equations (taking the limit $\Delta t\rightarrow0$) as
\begin{equation}
	\begin{aligned}
		\dot{x} &=\cos \varphi  & 	\dot{y}&=\sin \varphi\\
		\dot{\varphi} &= A_{0}\, \dfrac{L-R}{L+R +T_0}. &&
	\end{aligned}
	\label{eq:eomkernel}
\end{equation}
Here we assumed (again following~\cite{Perna2012}) that the ants' speed does not vary strongly and that we can choose the magnitude of the velocity to be constant, $v\equiv 1$~\footnote{Throughout the manuscript we assume nondimensionalization of all quantities and equations, where we have chosen system size and speed to be unity.}.

To transform the integro-differential equations of motion into ordinary differential equations, we Taylor-expand the pheromone density to first order
\begin{equation*}
	c(x+\Delta x,y+\Delta y)=c(x,y)+\partial_x c(x,y) \Delta x + \partial_y c(x,y) \Delta y,
\end{equation*}
and insert this into Eqs.~\ref{eq:kernalints1} and \ref{eq:kernalints2}.
In a coordinate system where $x$ is aligned to the velocity of the ant we find 
\begin{align*}
	L-R &=\dfrac{\sqrt{\pi}}{2}\;\mathcal{R}\; \partial_y c(x,y) \\
	L+R &=c(x,y)+\dfrac{\sqrt{\pi}}{2}\;\mathcal{R}\; \partial_x c(x,y).
\end{align*}
We see that the sensitivity of the ants rotation $A_0=A_0(\mathcal{R})$ in reaction to variation in the pheromones field needs to be inversely proportional to the size $\mathcal{R}$ of the detection domain, in order to keep the responses comparable. Writing 
$A_0(\mathcal{R})=A_0(\mathcal{R}_0) \mathcal{R}_0/\mathcal{R}$, with a arbitrarily chosen but sufficiently small $\mathcal{R}_0$,  we define
\begin{equation}
	\alpha=\lim_{\mathcal{R}\rightarrow 0}  \dfrac{\sqrt{\pi}}{2}\;\mathcal{R}\; A_0(\mathcal{R})=\dfrac{\sqrt{\pi}}{2} A_0(\mathcal{R}_0) \mathcal{R}_0,
\end{equation}
and after transforming back into the original (rotated) coordinate system, we find the equations of motion Eq.~\ref{eq:eom}.

\section{Random pheromone field model}
\label{app:rf}
\noindent
We generate a random (non-negative) concentration field $c(\vec{r})$ with prescribed correlation length $\ell_c$ by convolving $N$ randomly placed $\delta$-functions (in an area $\mathcal{L}\times \mathcal{L}$) with a Gaussian of width $\ell_g=\ell_c/\sqrt{2}$:

\[
c(\vec{r})=w\;g(\vec{r})\ast\sum_{i=1}^N\delta(\vec{r}_{i})
\]
with $g(\vec{r})={2 e^{- 2(x^{2}+y^{2})/\ell_c^{2}}}/\left( {\pi\ell_c^{2}}\right)$ and a global weight prefactor $w$. 
The mean density of $c(\vec{r})$ is $<c>=c_{o}=w\;N/\mathcal{L}^2={w\;n}/{\ell_c^{2}}$
where $n$ is the number of delta-functions per area $\ell_c^{2}$, i.e.\ $n=N \ell_c^2/\mathcal{L}^2$. For the correlation function we find
\begin{eqnarray*}
	C(\vec{r})&=&<c(\vec{r}'+\vec{r})c(\vec{r}')>-c_{0}^{2}\\
	&=&\frac{w^2\;n}{\pi \ell_c^4} e^{-r^2/\ell_c^2}.
\end{eqnarray*}
Throughout this article we will set $w=1$.


\section{Scaling of the characteristic length}
\label{app:scaling}
\noindent
To assess the characteristic length scale of branched flows of ant trajectories, we will follow Refs.~\cite{Kaplan2002,Degueldre2016,Degueldre2015} in using a simple scaling argument to find the parameter dependence of the \textit{mean distance to the first caustic} ($d_c$).
We recapitulate the scaling argument for initially parallel trajectory-bundles, i.e.\ the ant trajectory equivalent of an initially plane wave propagating through a correlated random, weakly refractive medium. This case is easier to understand and numerically simpler to test than the case of a point source, but (except for a different constant prefactor) yields the same results.

In the following we will assume that the sensitivity in the change of direction is sufficiently weak such that  $d_c \gg \ell_c$  and the mean free path ($\ell_\mathrm{mfp}$), i.e. the distance at which trajectories start to turn around, is much larger, $\ell_\mathrm{mfp}\gg d_c$.
We can then neglect the change in velocity in propagation direction (which we choose to be the $x$-direction), i.e.\ do a \textit{paraxial approximation}.

The main idea of the simple scaling argument is as follows: when a trajectory reaches a caustic (see red dots in Fig.~\ref{fig:pw1}c) the manifold connecting it with its neighbours in the bundle has a vertical tangent in phase space (therefore its projection onto coordinate space, here the y-axis, shows a singularity in the trajectory density). From the s-like shape of the manifold follows, however, that at the same y-position a second trajectory has to cross the first with different velocity (grey dot in in Fig.~\ref{fig:pw1}c), i.e. the trajectories cross under an angle that is finite (not infinitesimally small). Since the concentration field is correlated, however, for this to happen the trajectories need to have sufficiently different histories, i.e. they have to have initial conditions, that were on the order of a correlation length ($\ell_c$) apart. This initial distance has to have been spanned by diffusion in the random pheromone field. This is the origin of Eq.~\ref{eq:scalingArg}.

We thus have to study how the trajectories diffuse in the random pheromone field. If we follow a single trajectory its perpendicular velocity ($v_y$) will grow diffusively on timescales greater than $\ell_c/v_x \approx \ell_c/1$. We can approximate its dynamics by
\begin{align}
	\dot{y}&= v_y \\
	\dot{v}_y &=  \frac{\alpha \partial_y c}{c+T_0} \approx \gamma_p\; \Gamma(t),\label{eq:vydiff}
\end{align}
 with $\left\langle \Gamma(t')\Gamma(t)\right\rangle =\delta(t'-t)$. The prefactor $\gamma_p=\sqrt{2 D_p}$ determines the diffusive growth : $v_y= 2 D_p t$ caused by the random (but static) pheromone field.
 It can by found by directly integrating Eq.~\ref{eq:vydiff} to be $\gamma_p=\alpha\;r_c$, with the \textit{correlation radius} $r_c$ of the fluctuating force Eq.~\ref{eq:corrrad}.

The variance in $y$ can be found to be (see e.g.~\cite{Degueldre2015})
\begin{equation}
	\left\langle (y-y_0)^2 \right\rangle = \frac{2}{3} D_p t^3.
\end{equation}
From inserting this into Eq.~\ref{eq:scalingArg} the power-law of Eq.~\ref{eq:dcscaling} easily follows (since $x\approx t$). 

Note, that when the different scales are less well separated ($\ell_\mathrm{mfp} \gtrsim  d_c \gtrsim  \ell_c$), the paraxial approximation will be less accurate and deviations from the scaling law will occur. However, the phenomenon of branched flow can still be observed.

\section{Stability matrix and caustic condition}
\label{app:stabMat}
\noindent
To efficiently calculate the caustic statistics of branched flows in the ant-dynamics, we follow the methods developed in Ref.~\cite{Metzger2010a} and numerically evaluate the stability matrix along the trajectories until we reach a \textit{caustic condition}. 
The stability matrix $\mat{M}(t,t_0)$ describes how a trajectory $\vec{x}(t) = \vec{x}_0(t) + \delta\vec{x}(t)$, that starts infinitesimally close to a reference trajectory $\vec{x}_0(t)$ at time $t=t_0$ evolves over time (here $\vec{x}(t) = (x(t),y(t),\varphi(t))^T$), i.e.
\begin{equation*}
	\delta \vec{x}(t) = \mat{M}\;\delta\vec{x}(t_0).
\end{equation*}
The elements of $\mat{M}$ are $M_{ij}(t,t_0)=\partial x_i(t)/\partial x_j(t_0)$.
Their dynamics is given by
\begin{equation}
	\dot{\mat{M}}=\mat{K}\;\mat{M},
\end{equation}
with the Jacobian matrix of the equations of motion $K_{ij}(t)=\partial \dot{x}_i(t)/\partial x_j(t)$,
i.e.
\begin{equation}
	\mat{K}  = \left( \begin{matrix} 
		0 & 0 & -\sin \varphi \\ 
		0 & 0 & \cos \varphi \\ 
		K_{31} & K_{32} & K_{33} 
	\end{matrix} \right),
\end{equation}
with 
\begin{align*}
K_{31} &= \tfrac{A(- c_{xx} \sin \varphi + c_{xy} \cos \varphi)}{T_0+c} - \tfrac{A c_x (-c_x \sin \varphi + c_y \cos \varphi)}{(T_0+c)^2},\\
K_{32} &= \tfrac{A(c_{yy} \cos \varphi -c_{xy} \sin \varphi)}{T_0+c} - \tfrac{A c_y (-c_x \sin \varphi + c_y \cos \varphi)}{(T_0+c)^2},\\
K_{33} &= \tfrac{A(-c_y \sin \varphi - c_x \cos \varphi)}{T_0 + c}
\end{align*}
where $c_{i}=\partial c/\partial x_i$ and $c_{ij}=\partial^2 c/\partial x_i \partial x_j$.

The reference trajectory reaches a caustic, i.e.\ a singularity in the trajectory density, when the area of the projection onto coordinate space of the parallelogram spanned in phase space by $\dot{\vec{x}}(t)$ and $\delta\vec{x}(t)$ vanishes. For initially parallel trajectories ($\dot{\vec{x}}(t_0)=(1,0,0)^T$ and $\delta \vec{x}(t_0)=(0,1,0)^T$) we can write the caustic condition as 
\begin{equation}
	\dot{x}\;M_{22} - \dot{y}\;M_{12}=M_{22}\cos \varphi - M_{12} \sin \varphi=0.
\end{equation}

\section{Supplemental Material}
\label{app:videos}

\noindent
\href{https://arxiv.org/src/2308.11232v2/anc/Video_1-Example_rotational_diffusion.mp4}{\textbf{Video 1}: \textbf{Example of a flow of ant trajectories} from a point source in a random pheromone field with increasing stochastic rotational diffusion (growing $\gamma_2$). The other parameters are $\alpha=0.2$, $T_0=0.0$, $\ell_c=0.005$, and $n=1.0$.}

\vspace{\baselineskip}
\noindent
\href{https://arxiv.org/src/2308.11232v2/anc/Video_2-Trailpattern_evolution.mp4}{\textbf{Video 2}: \textbf{Example of the trailpattern evolution} as described in Figs.~\ref{fig:trails} and \ref{fig:phasespace}. The left panel of the \textbf{upper row} shows the accumulated pheromone field successively deposited by the ants. The centre panel shows the latest $10$, respectively $100$ ant trajectories. The right panel reveals the density (on a logarithmic scale) of \textit{potential trajectories} of the next ant in the current pheromone field. The right panel in the \textbf{lower row} shows the density of \textit{potential trajectories} in the absence of stochastic diffusion terms. The left panel shows Poincaré surfaces of section along the circles of corresponding colour indicated in the right panel. }

\bibliography{../../References}

\end{document}